\documentclass[aps,prb,citeautoscript,twocolumn,superscriptaddress,footinbib,10pt]{revtex4-1} 
\usepackage[latin1]{inputenc}
\usepackage{amsmath,amssymb}
\usepackage{mathrsfs} 
\usepackage{graphicx,color,colortbl}
\usepackage{rotating,array,tabularx,booktabs}
\usepackage{varwidth,xcolor}
\usepackage{placeins}
\usepackage{siunitx}

\DeclareGraphicsExtensions{.pdf,.PDF,.png,.PNG}

\newcommand{\dif}{\mathrm{d}}%
\newcommand{\pdif}[2]{\frac{\partial#1}{\partial#2}}%
\newcommand{\ZT}[1]{\textquotedblleft#1\textquotedblright}%

\begin{document}

\title{On the shape-dependent propulsion of nano- and microparticles by traveling ultrasound waves}

\author{Johannes Vo\ss{}}
\affiliation{Institut f\"ur Theoretische Physik, Center for Soft Nanoscience, Westf\"alische Wilhelms-Universit\"at M\"unster, D-48149 M\"unster, Germany}

\author{Raphael Wittkowski}
\email[Corresponding author: ]{raphael.wittkowski@uni-muenster.de}
\affiliation{Institut f\"ur Theoretische Physik, Center for Soft Nanoscience, Westf\"alische Wilhelms-Universit\"at M\"unster, D-48149 M\"unster, Germany}

\begin{abstract}
Among the many types of artificial motile nano- and microparticles that have been developed in the past, colloidal particles that exhibit propulsion when they are exposed to ultrasound are particularly advantageous. Their properties, however, are still largely unexplored. For example, the dependence of the propulsion on the particle shape and the structure of the flow field generated around the particles are still unknown. In this article, we address the propulsion mechanism of ultrasound-propelled nano- and microparticles in more detail. Based on direct computational fluid dynamics simulations and focusing on traveling ultrasound waves, we study the effect of two important aspects of the particle shape on the propulsion: rounded vs.\ pointed and filled vs.\ hollow shapes. We also address the flow field generated around such particles. Our results reveal that pointedness leads to an increase of the propulsion speed, whereas it is not significantly affected by hollowness. Furthermore, we find that the flow field of ultrasound-propelled particles allows to classify them as pusher squirmers, which has far-reaching consequences for the understanding of these particles and allows us to predict that they can be used to realize active materials with a tunable viscosity that can exhibit suprafluidity and even negative viscosities. The obtained results are helpful, e.g., for future experimental work further investigating or applying ultrasound-propelled colloidal particles as well as for theoretical approaches that aim at modeling their dynamics on mesoscopic scales.
\begin{figure}[htb]
\centering
\fbox{\includegraphics[width=8cm]{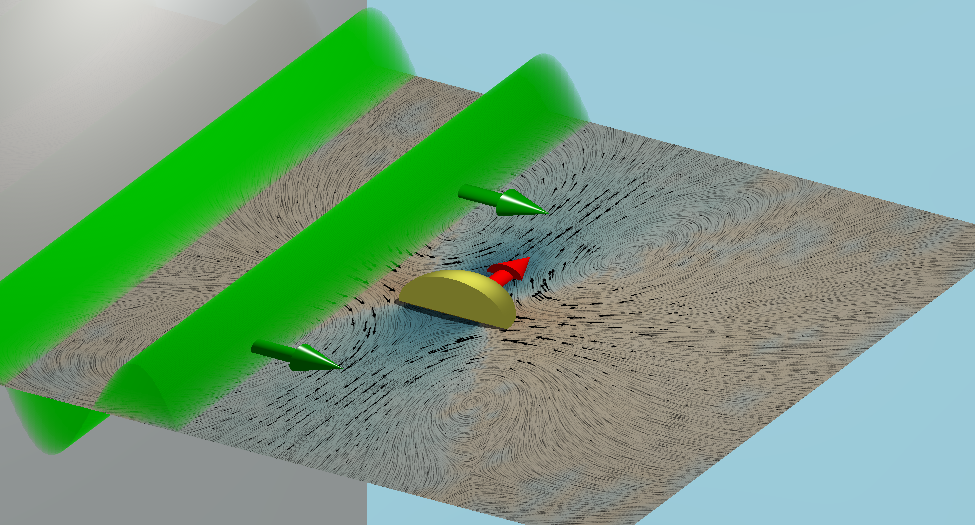}}%
\end{figure}
\end{abstract}
\maketitle

\section{Introduction}
With the experimental discovery of fuel-less ultrasound-propelled (also called \ZT{self-acoustophoretic}) colloidal particles in 2012 \cite{WangCHM2012}, important potential applications of motile nano- and microparticles (also called \ZT{active particles})\cite{BechingerdLLRVV2016} have become within reach \cite{XuXZ2017}. One of their most attractive potential applications is the usage as self-propelled nano- or microdevices in medicine \cite{KimGLF2018,WuEtAl2014,WuEtAl2015b} that allows for targeted drug delivery \cite{GarciaGradillaSSKYWGW2014,EstebanFernandezEtAl2017,UygunEtAl2017,LuoFWG2018,WangLMAHM2014,EstebanEtAl2016}, enhanced biodetoxification \cite{WuEtAl2015a,EstebanEtAl2018}, nanosurgery \cite{LiEFdAGZW2017,ErkocYCYAS2019}, enhanced diagnostics \cite{ChalupniakEtAl2015,KimGLF2018,QualliotineEtAl2019,EstebanFernandezdeAvilaMSLRCVMGZW2015}, and many more fascinating applications \cite{ErkocYCYAS2019,PengTW2017,SotoC2018,HeFHLGH2016,BalasubramanianEtAl2011,GaoEFdAZW2018,HuHZHZJZ2018,XuGXZW2017,WangGWSGXH2018}. In contrast to different propulsion mechanisms like chemical propulsion \cite{EstebanFernandezdeAvilaALGZW2018,SafdarKJ2018,PengTW2017}, other fuel-based propulsion \cite{KaganBCCEEW2012}, light-propulsion \cite{XuanSGWDH2018}, and X-Ray propulsion \cite{XuCLFPLK2019}, the acoustic propulsion \cite{WangCHM2012,GarciaGradillaEtAl2013,AhmedEtAl2013,WuEtAl2014,GarciaGradillaSSKYWGW2014,BalkEtAl2014,AhmedGFM2014,WangDZSSM2015,WuEtAl2015a,WuEtAl2015b,SotoWGGGLKACW2016,AhmedWBGHM2016,UygunEtAl2017,EstebanFernandezEtAl2017,RenZMXHM2017,SabrinaTABdlCMB2018,AhmedBJPDN2016,LuSZWPL2019,TangEtAl2019,QualliotineEtAl2019} 
has important advantages: It is fuel-free, biocompatible, and allows to supply the particles continuously with energy. 
Ultrasound-propelled particles can be rigid \cite{WangCHM2012,GarciaGradillaEtAl2013,AhmedEtAl2013,NadalL2014,BalkEtAl2014,AhmedGFM2014,GarciaGradillaSSKYWGW2014,WangDZSSM2015,SotoWGGGLKACW2016,AhmedBJPDN2016,UygunEtAl2017,CollisCS2017,SabrinaTABdlCMB2018,TangEtAl2019,ZhouZWW2017} or have moveable components \cite{KaganBCCEEW2012,AhmedLNLSMCH2015,AhmedBJPDN2016,RenEtAl2019}. The latter ones include bubble-propelled particles \cite{KaganBCCEEW2012,RenEtAl2019}, which can reach rather high propulsion speeds, but since the former ones are easier to produce, they are more likely to be applied in the near future. There exist also hybrid particles combining acoustic propulsion with a different propulsion mechanism \cite{LILXKLWW2015,WangDZSSM2015,RenZMXHM2017,TangEtAl2019,RenWM2018}. 

Ultrasound-propelled nano- and microparticles have been intensively investigated in recent years
\cite{WangCHM2012,GarciaGradillaEtAl2013,AhmedEtAl2013,BalkEtAl2014,NadalL2014,SotoWGGGLKACW2016,AhmedWBGHM2016,AhmedBJPDN2016,CollisCS2017,SabrinaTABdlCMB2018,TangEtAl2019,RaoLMZCW2015,KimGLZF2016,ZhouZWW2017,ChenEtAl2018}. While the most studies are based on experiments \cite{WangCHM2012,GarciaGradillaEtAl2013,AhmedEtAl2013,BalkEtAl2014,SotoWGGGLKACW2016,AhmedWBGHM2016,AhmedBJPDN2016,SabrinaTABdlCMB2018,TangEtAl2019,ZhouZWW2017}, there are only two theory-based studies so far \cite{NadalL2014,CollisCS2017}. 
Despite the large number of existing studies on acoustically propelled particles, we are still at the beginning of exploring and understanding their features. 
Even the details of their propulsion mechanism are still unclear. 
For example, it is not yet known how the propulsion speed depends on the properties of the particles and their environment, what the maximal speed of the particles for a given ultrasound intensity is, and which structure the flow field generated around the particles has. One of the most basic properties of the particles is their shape. Nevertheless, only very few particle shapes have been considered so far.  
The main reason for this is that the particle shape cannot easily be varied in experimental studies and that the existing theoretical studies focus on the particle shapes used in the experiments. 
The particle shapes studied so far are mostly cylinders with a concave and a convex end  \cite{WangCHM2012,AhmedEtAl2013,AhmedGFM2014,BalkEtAl2014,ZhouZWW2017}. 
As a limiting case, also cup-shaped particles were studied \cite{SotoWGGGLKACW2016,TangEtAl2019}. 
Apart from that, there exist only a study that addresses gear-shaped particles \cite{SabrinaTABdlCMB2018} and studies on particles with movable components \cite{KaganBCCEEW2012,AhmedLNLSMCH2015,AhmedBJPDN2016,RenEtAl2019} and thus a nonconstant shape.

For a cylindrical shape with spherical concave and convex ends, the direction of movement was found in experiments to point towards the concave end \cite{AhmedWBGHM2016}, but the two theoretical studies suggest that this depends on the shape of the caps \cite{NadalL2014,CollisCS2017}. The direction of propulsion seems to depend also on the length of the cylinder, since cup-shaped particles were found to move towards their convex end \cite{SotoWGGGLKACW2016}.
Hence, we can conclude that the propulsion speed depends sensitively on the particles' shape, but we have not yet a deeper understanding of this dependence.  

A better understanding of the shape-dependence of the propulsion would be helpful for future studies, since it would provide a good opportunity for optimizing the particle speed and thus the efficiency of the particles' propulsion. 
Large propulsion speeds are crucial for medical applications, where the maximal ultrasound intensity is limited by the requirement of biocompatibility and the particles must be fast enough to withstand the blood flow that tends to carry the particles away. 
The fastest ultrasound-propelled fuel-free particles observed so far reached a speed of about $\SI{250}{\micro\meter\,\second^{-1}}$ \cite{GarciaGradillaEtAl2013}. This is faster than the blood flow in the vascular capillaries with a typical speed of about $\SI{100}{\micro \meter \,\second^{-1}}$ \cite{ErkocYCYAS2019}, but the transducer voltage of $\SI{10}{\volt}$ applied in the corresponding experiments indicates that the acoustic energy density was about $\SI{10}{}$-$\SI{100}{\joule\,\metre^{-3}}$ \cite{Bruus2012} and thus too high for usage in the human body, where the energy density should be below $\SI{4.9}{\joule\,\metre^{-3}}$ to avoid damage to the tissue \cite{QBarnettEtAl2000}. 

A further limitation related to the experiments is the usage of a standing ultrasound wave in all but one  \cite{AhmedBJPDN2016} experimental studies.
To facilitate observation of the particles with a microscope, they are enclosed by a thin chamber with two parallel horizontal walls, of which at least the upper one is transparent, and a standing wave field that levitates the particles in a nodal plane between the horizontal walls of the chamber.
However, in many important potential applications, such as medical ones, the sound waves would be traveling. 

In this article, we advance the knowledge about the acoustic propulsion of homogeneous rigid nano- and microparticles that are exposed to traveling ultrasound waves. Using direct computational fluid dynamics simulations based on numerically solving the compressible Navier-Stokes equations, we determine the sound-induced forces acting on these particles together with their resulting propulsion speed as well as the flow field generated around the particles.
To address the shape-dependence of these quantities, we consider some particle shapes that differ with respect to two aspects not studied previously: We compare rounded with pointed shapes and filled with hollow ones.

\section{\label{methods}Methods}
\begin{figure}[htb]
\centering
\includegraphics[width=\linewidth]{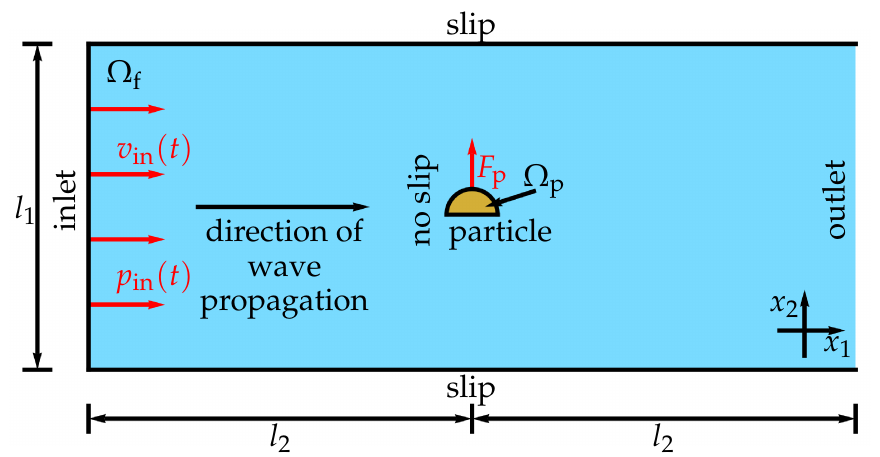}%
\caption{\label{fig:setup}Setup for the simulations. A traveling ultrasound wave is entering the fluid domain $\Omega_{\mathrm{f}}$ at the inlet, where it is prescribed by an inflow velocity $v_{\mathrm{in}}(t)$ and pressure $p_{\mathrm{in}}(t)$. The width of the fluid domain is $l_1$ and the rigid particle, constituting a particle domain $\Omega_{\mathrm{p}}$, is placed at a distance $l_2$ from the inlet. At the particle boundary a no-slip condition is prescribed and for the lateral boundaries of $\Omega_{\mathrm{f}}$ slip boundary conditions are used. The ultrasound exerts a time-averaged propulsion force $F_{\mathrm{p}}$ on the particle and after a further distance $l_2$ the domain $\Omega_{\mathrm{f}}$ ends with an outlet.}
\end{figure}

The setup for our study is shown in Fig.\ \ref{fig:setup}.
A traveling ultrasound wave with frequency $f=\SI{1}{\MHz}$ enters a rectangular domain $\Omega_{\mathrm{f}}$ of an initially quiescent fluid, which we assume to be water, at an inlet of width $l_1=200\sigma$, where $\sigma=\SI{1}{\micro\metre}$ is the particle diameter. At the inlet, the ultrasound wave is prescribed by a time-dependent inflow velocity $v_{\mathrm{in}}(t)=(\Delta p / (\rho_0 c_{\mathrm{f}})) \sin(2\pi f t)$ perpendicular to the inlet and a time-dependent inflow pressure $p_{\mathrm{in}}(t)=\Delta p \sin(2\pi f t)$ with the pressure amplitude $\Delta p=\SI{10}{\kilo \pascal}$, the density of the quiescent fluid $\rho_0=\SI{998}{\kilogram\,\metre^{-3}}$, and the sound velocity in the fluid $c_{\mathrm{f}}=\SI{1484}{\metre\,\second^{-1}}$. The pressure amplitude corresponds to an acoustic energy density $E=\Delta p^2/(2 \rho_0 c_{\mathrm{f}}^2)=\SI{22.7}{\milli\joule\,\metre^{-3}}$. Starting at the inlet, the ultrasound wave propagates through the fluid domain parallel to its lateral boundaries, where we prescribe slip boundary conditions. After a distance $l_2=\lambda/4$ with the wavelength of the ultrasound $\lambda=\SI{1.484}{\milli\metre}$, the wave arrives the fixed rigid particle, which is oriented perpendicular to the propagation direction of the wave. In the simulations, we describe the particle by a particle domain $\Omega_{\mathrm{p}}$ and prescribe no-slip conditions at the particle's boundary $\partial\Omega_{\mathrm{p}}$. Through the ultrasound, a propulsion force $F_{\mathrm{p}}$ parallel to the particle's orientation is exerted on the particle. The wave then propagates a further distance $l_2$ until it reaches an outlet at the end of the domain $\Omega_{\mathrm{f}}$.

To avoid approximations like perturbation expansions that are involved in all previous studies using analytical \cite{NadalL2014,CollisCS2017} or numerical \cite{AhmedBJPDN2016,SabrinaTABdlCMB2018,TangEtAl2019} methods to determine the propulsion speed of acoustically propelled particles, we base our work on direct fluid dynamics simulations. Our simulations are carried out by numerically solving the compressible Navier-Stokes equations together with the continuity equation for the mass-density field $\rho(\vec{x},t)$ of the fluid and a constitutive equation for the fluid's pressure field $p(\vec{x},t)$ in the two-dimensional fluid domain $\Omega_{\mathrm{f}}$. Here, $\vec{x}=(x_{1},x_{2})^{\mathrm{T}}$ is the position vector and $t$ denotes time. 

When $\vec{v}(\vec{x},t)$ with $\vec{v}=(v_{1},v_{2})^{\mathrm{T}}$ is the velocity field of the fluid and we use the short notation $\partial_{i}=\partial/\partial_{x_{i}}$ with $i\in\{1,2\}$ for the spatial derivatives, the \textit{continuity equation} that describes mass conservation is given by \cite{LandauL1987}
\begin{equation}
\pdif{\rho}{t} + \sum^{2}_{i=1}\partial_{i}(\rho v_{i}) = 0.
\label{eq:Massenerhaltung}%
\end{equation}
Momentum conservation of the fluid is then described by the \textit{Navier-Stokes equations} \cite{LandauL1987}
\begin{equation}
\pdif{}{t}(\rho v_{i}) + \sum^{2}_{j=1}\partial_{j}\Pi_{ij} = 0
\label{eq:Impulserhaltung}%
\end{equation}
with the momentum-current tensor
\begin{equation}
\Pi_{ij} = \rho v_{i}v_{j} - \Sigma_{ij}
\label{eq:Impulsstrom}%
\end{equation}
and the stress tensor
\begin{equation}
\Sigma_{ij} = \Sigma^{(\mathrm{p})}_{ij}+\Sigma^{(\mathrm{v})}_{ij}
\label{eq:Spannungstensor}%
\end{equation}
for $i,j\in\{1,2\}$. The stress tensor consists of the pressure part 
\begin{equation}
\Sigma^{(\mathrm{p})}_{ij} = -p\delta_{ij}
\label{eq:Spannungstensorp}%
\end{equation}
and viscous part
\begin{equation}
\begin{split}%
\Sigma^{(\mathrm{v})}_{ij} &= \nu_{\mathrm{s}}\bigg( \partial_{j}v_{i} + \partial_{i}v_{j} - \frac{2}{3}\sum^{2}_{k=1}(\partial_{k}v_{k})\delta_{ij} \bigg)\\
&\quad\:\! + \nu_{\mathrm{b}}\sum^{2}_{k=1}(\partial_{k}v_{k})\delta_{ij}
\end{split}%
\label{eq:Spannungstensorv}%
\end{equation}
with the Kronecker symbol $\delta_{ij}$, shear viscosity (also called \ZT{dynamic viscosity}) $\nu_{\mathrm{s}}$, and bulk viscosity (also called \ZT{volume viscosity}) $\nu_{\mathrm{b}}$. Heat conduction and heating of the fluid by the sound waves are neglected here, since the ultrasound intensities that are used in experiments with acoustically-propelled colloidal particles are usually rather small.
To close the set of Eqs.\ \eqref{eq:Massenerhaltung}-\eqref{eq:Spannungstensorv}, we need a constitutive equation for the pressure $p(\vec{x},t)$.
When the sound intensity is sufficiently small, so that the fluid is acoustically nondispersive and heating of the fluid by the sound wave can be neglected, the local pressure $p(\vec{x},t)$ is given by the \textit{constitutive equation}
\begin{equation}
p(\rho) = p_{0} + c_{\mathrm{f}}^{2} (\rho - \rho_{0})
\label{eq:Druck}%
\end{equation}
as a function of the local mass density $\rho(\vec{x},t)$.
Here, $c_{\mathrm{f}}$ is the speed of sound in the fluid and $\rho_{0}$ and $p_{0}=p(\rho_{0})$ are the constant mean mass density and pressure of the fluid, respectively.
To solve Eqs.\ \eqref{eq:Massenerhaltung}-\eqref{eq:Druck} numerically, we used the software package OpenFOAM \cite{WellerTJF1998}, which applies the finite volume method.

The time-dependent force $\vec{F}_{\mathrm{p}}(t)$ with $\vec{F}_{\mathrm{p}}=(F_{\mathrm{p},1},F_{\mathrm{p},2})^{\mathrm{T}}$ acting on the particle is calculated in the laboratory frame.
Since the particle, which is described by the particle domain $\Omega_{\mathrm{p}}$, has no-slip boundary conditions and is fixed in space in our simulations, the fluid velocity $\vec{v}(\vec{x},t)$ is zero at the fluid-particle interface $\partial\Omega_\mathrm{p}$. So the force acting on the particle is given by $\vec{F}_{\mathrm{p}}=\vec{F}^{(\mathrm{p})}+\vec{F}^{(\mathrm{v})}$ with the components \cite{LandauL1987}
{\allowdisplaybreaks\begin{align}%
F^{(\alpha)}_{i} &=  \sum^{2}_{j=1} \int_{\partial\Omega_{\mathrm{p}}} \!\!\!\!\!\!\! \Sigma^{(\alpha)}_{ij}\,\dif A_{j}, \qquad \alpha\in\{\mathrm{p},\mathrm{v}\}.
\label{eq:F}%
\end{align}}%
Here, $\dif\vec{A}(\vec{x})$ with $\dif\vec{A}=(\dif A_{1},\dif A_{2})^{\mathrm{T}}$ is the normal and outwards oriented surface element of $\partial\Omega_{\mathrm{p}}$ at position $\vec{x}$ when $\vec{x}\in\partial\Omega_{\mathrm{p}}$.
By time-averaging $\vec{F}^{(\mathrm{p})}(t)$ and $\vec{F}^{(\mathrm{v})}(t)$ locally over one period $\tau$ of the ultrasound wave in the stationary state (i.e., for large $t$), we obtain the time-averaged stationary forces $\langle\vec{F}^{(\mathrm{p})}\rangle$, $\langle\vec{F}^{(\mathrm{v})}\rangle$, and $\langle\vec{F}_{\mathrm{p}}\rangle=\langle\vec{F}^{(\mathrm{p})}\rangle+\langle\vec{F}^{(\mathrm{v})}\rangle$, where $\langle\cdot\rangle$ denotes the time average.
Using $\langle\vec{F}_{\mathrm{p}}\rangle$, we then calculate the time-averaged stationary propulsion velocity \cite{HappelB1991,VossW2018}
\begin{equation}
\langle\vec{v}_{\mathrm{p}}\rangle = \frac{1}{\nu_\mathrm{s}} \boldsymbol{\mathrm{K}}^{-1} \langle\vec{F}_{\mathrm{p}}\rangle 
\label{eq:resistance}%
\end{equation}
with the resistance matrix $\boldsymbol{\mathrm{K}}$ of the considered particle, which is determined using the software \texttt{HydResMat} \cite{VossW2018,VossJW2019}.
As we simulate a two-dimensional system to keep the computational effort manageable, but $\boldsymbol{\mathrm{K}}$ is a $3\times 3$-dimensional matrix that corresponds to a three-dimensional particle, we cannot apply Eqs.\ \eqref{eq:F} and \eqref{eq:resistance} directly. Therefore, we assign a thickness of $\sigma$ to the particle, which equals its diameter, so that $\boldsymbol{\mathrm{K}}$ can be calculated. 
Neglecting contributions by the lower and upper surfaces of the particle, we then use the three-dimensional versions of Eqs.\ \eqref{eq:F} and \eqref{eq:resistance}. 
From $\langle\vec{F}_{\mathrm{p}}\rangle$ and $\langle\vec{v}_{\mathrm{p}}\rangle$, we directly obtain the propulsion force $F_{\mathrm{p}}=(\langle\vec{F}_{\mathrm{p}}\rangle)_2=F_{\mathrm{p},p}+F_{\mathrm{p},v}$, its pressure component $F_{\mathrm{p},p}=(\langle\vec{F}^{\mathrm{(p)}}\rangle)_2$ and viscous component $F_{\mathrm{p},v}=(\langle\vec{F}^{\mathrm{(v)}}\rangle)_2$, as well as the propulsion speed $v_{\mathrm{p}}=(\langle\vec{v}_{\mathrm{p}}\rangle)_2$ as the force- and velocity contributions parallel to the particle's orientation, i.e., parallel to the $x_2$ axis. 

Nondimensionalization of the equations introduced above leads to the Helmholtz number $\mathrm{He}$, a Reynolds number corresponding to the shear viscosity $\mathrm{Re}_\mathrm{s}$, another Reynolds number corresponding to the bulk viscosity $\mathrm{Re}_\mathrm{b}$, and the product $\mathrm{Ma}^2\mathrm{Eu}$, with the Mach number $\mathrm{Ma}$ and Euler number $\mathrm{Eu}$, corresponding to the pressure amplitude $\Delta p$ of the ultrasound wave that enters the simulated system. 
Table \ref{tab:Parameters} shows the names, symbols, and assigned values of the parameters that are relevant for our simulations.
\begin{table}[htb]
\centering
\begin{ruledtabular}
\begin{tabular}{p{46mm}ccc}%
\textbf{Name} & \textbf{Symbol} & \textbf{Value}\\
\hline
Particle diameter & $\sigma$ & $\SI{1}{\micro\metre}$\\
Sound frequency & $f$ & $\SI{1}{\mega\hertz}$\\
Speed of sound & $c_\mathrm{f}$ & $\SI{1484}{\metre\,\second^{-1}}$\\
Time period of sound & $\tau=1/f$ & $\SI{1}{\micro\second}$\\
Wavelength of sound & $\lambda=c_\mathrm{f}/f$ & $\SI{1.484}{\milli\metre}$\\
Mean mass density of fluid & $\rho_0$ & $\SI{998}{\kilogram\,\metre^{-3}}$\\
Mean pressure of fluid & $p_{0}$ & $\SI{101325}{\pascal}$ \\
Initial velocity of fluid & $\vec{v}_{0}$ & $\vec{0}\,\SI{}{\metre\,\second^{-1}}$ \\
Sound pressure amplitude & $\Delta p$ & \SI{10}{\kilo\pascal}\\
Acoustic energy density & \hspace*{-4mm}$E=\Delta p^2/(2 \rho_0 c_{\mathrm{f}}^2)$\hspace*{-1mm} & $\SI{22.7}{\milli\joule\,m^{-3}}$\\
Shear/dynamic viscosity of fluid & $\nu_{\mathrm{s}}$ & $\SI{1.002}{\milli\pascal\,\second}$ \\
Bulk/volume viscosity of fluid & $\nu_{\mathrm{b}}$ & $\SI{2.87}{\milli\pascal\,\second}$ \\
Domain width & $l_1$ & $200\sigma$\\
\mbox{Inlet-particle or particle-outlet} distance & $l_2$ & $\lambda/4$ \\
Mesh-cell size & $\Delta x$ & $\SI{15}{\nano \metre}$-$\SI{1}{\micro \metre}$ \\
Time-step size & $\Delta t$ & $1$-$\SI{10}{\pico \second}$\\
Simulation duration & $t_{\mathrm{max}}$ & $500\tau$ \\
\end{tabular}%
\end{ruledtabular}%
\caption{\label{tab:Parameters}Parameters that are relevant for our simulations and their values. We obtained the bulk viscosity $\nu_\mathrm{b}$ for water at temperature $T=\SI{293.15}{\kelvin}$ by a cubic spline interpolation of the data from Tab.\ 1 in Ref.\ \onlinecite{HolmesPP2011}.}%
\end{table}
The parameters related to the fluid are based on assuming that the fluid is water at normal temperature $T=\SI{293.15}{\kelvin}$ and normal pressure $p_0=\SI{101325}{\pascal}$. 
With the parameter values from Tab.\ \ref{tab:Parameters}, our simulations correspond to the following values of the dimensionless numbers: 
\begin{align}
\mathrm{He}&=2 \pi f \sigma / c_\mathrm{f}\approx 4.234\cdot 10^{-3},\\
\mathrm{Re}_\mathrm{s}&=\rho_0 c_\mathrm{f} \sigma / \nu_\mathrm{s}\approx 1478,\\
\mathrm{Re}_\mathrm{b}&=\rho_0 c_\mathrm{f} \sigma / \nu_\mathrm{b}\approx 516,\\
\mathrm{Ma}^2 \mathrm{Eu}&=\Delta p/(\rho_0 c_\mathrm{f}^2)\approx 4.550\cdot 10^{-6}.
\end{align}

We discretized the fluid domain $\Omega_{\mathrm{f}}$ using a structured mixed rectangular-triangular mesh with about 250,000 cells. The typical cell size $\Delta x$ varied from about $\SI{15}{\nano\metre}$ near the particle to about $\SI{1}{\micro \metre}$ far away from the particle. For the time integration, we used an adaptive time-step method with a maximum time-step size ensuring that the Courant-Friedrichs-Lewy number 
\begin{align}
C = c_\mathrm{f} \frac{\Delta t}{\Delta x}
\end{align}
is smaller than one. The typical time-step size $\Delta t$ was thus between $\SI{1}{\pico\second}$ and $\SI{10}{\pico\second}$.
The simulations ran for $t_\mathrm{max}= 500\tau$ to get sufficiently close to the stationary state. 
Due to the fine discretization in space and time and the relatively large spatial and temporal domains, the simulations were computationally very expensive and required a strong parallelization. 
The typical duration of one simulation was about $36,000$ CPU core hours.

\begin{figure}[tb]
\centering
\includegraphics[width=\linewidth]{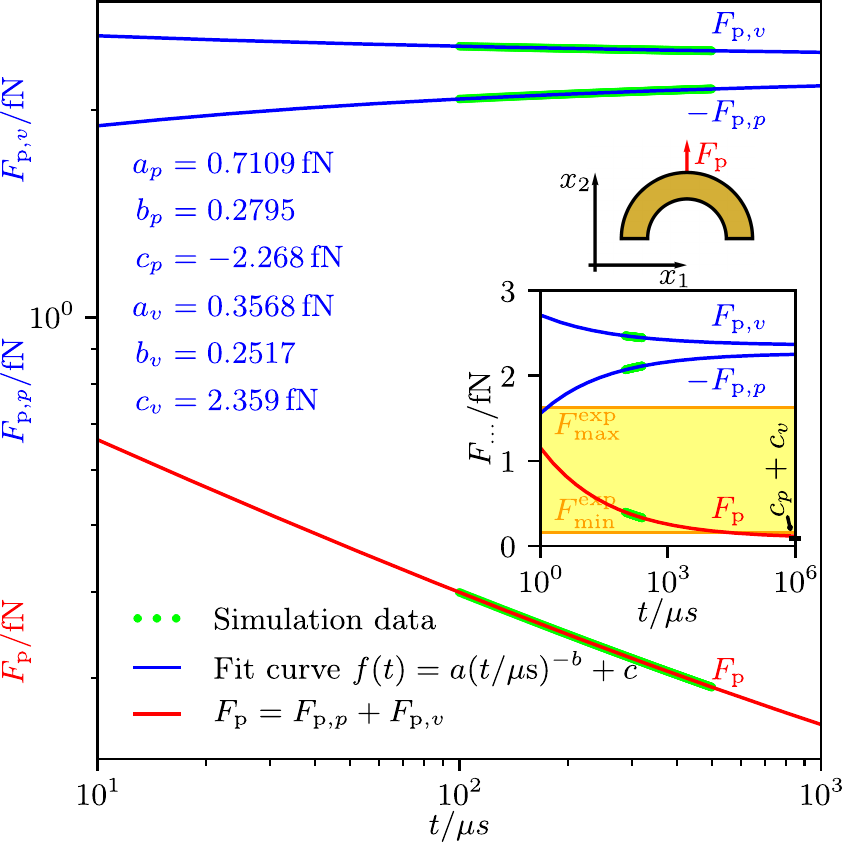}%
\caption{\label{fig:Fit_Plot}Simulation data for the time-dependent forces $F_{\mathrm{p},p}(t)$ and $F_{\mathrm{p},v}(t)$ acting on a particle with the shape of a hollow half ball as well as an extrapolation of the forces with the fit function $f(t)$. The extrapolation of the total propulsion force $F_{\mathrm{p}}(t)=F_{\mathrm{p},p}(t)+F_{\mathrm{p},v}(t)$ converges against $c_p+c_v$, where $c_p$ and $c_v$ are the offset fit coefficients in $f(t)$ for $F_{\mathrm{p},p}(t)$ and $F_{\mathrm{p},v}(t)$, respectively. Its limiting value is consistent with corresponding experimental data from Soto \textit{et al.} \cite{SotoWGGGLKACW2016} that can be tied to the interval $[F^{\mathrm{exp}}_{\mathrm{min}},F^{\mathrm{exp}}_{\mathrm{max}}]=[\SI{0.163}{\femto\newton},\SI{1.63}{\femto\newton}]$.}
\end{figure}

Since the simulations would require even more time to fully converge, we determined the stationary forces $F_{\mathrm{p},p}$ and $F_{\mathrm{p},v}$ by extrapolation. For this purpose, we used the fit function
\begin{equation}
f(t) = a \bigg(\frac{t}{\SI{}{\micro\second}}\bigg)^{-b} + c .
\label{eq:Fit}%
\end{equation}
An example for the extrapolation, corresponding to a hollow-half-ball particle, is shown in Fig.\ \ref{fig:Fit_Plot}. 
In the inset of this figure, the experimental data \cite{SotoWGGGLKACW2016} available for this particle shape are indicated by a yellow band.
The values of the fit parameters for all considered particle shapes are listed in Tab.\ \ref{tab:Fitparameter}.
\begin{table*}[tb]
\centering
\begin{ruledtabular}
\begin{tabular}{ccccc}
\textbf{Fit parameter} & \includegraphics[scale=0.085]{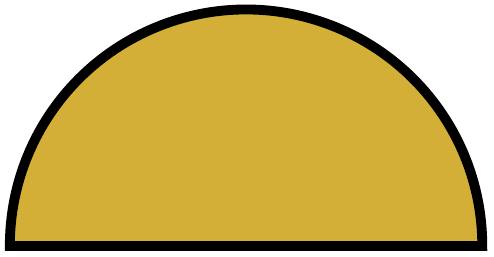}\;\textbf{Half ball} & \includegraphics[scale=0.085]{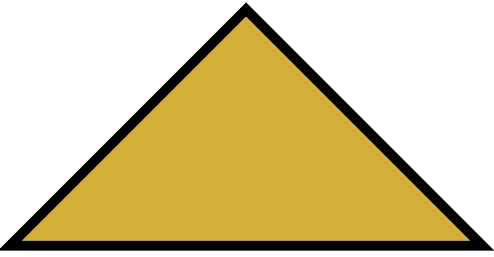}\;\textbf{Cone} & \includegraphics[scale=0.085]{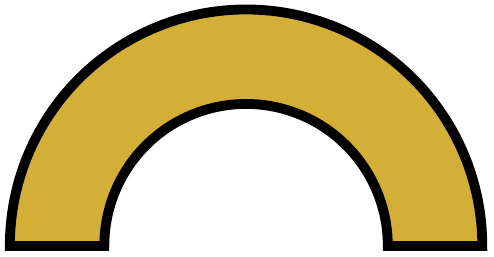}\;\textbf{Hollow half ball} & \includegraphics[scale=0.085]{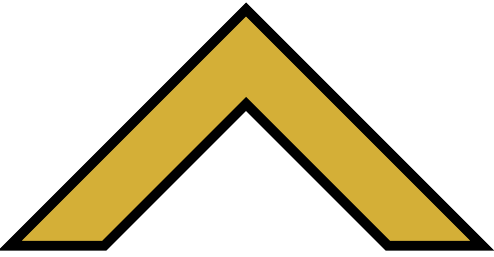}\;\textbf{Hollow cone} \\
\hline
$a_p$ & $\SI{0.5188}{\femto\newton}$ & $\SI{0.7301}{\femto\newton}$ & $\SI{0.7109}{\femto\newton}$ & $\SI{0.9332}{\femto\newton}$\\
$b_p$ & $\SI{0.2825}{}$ & $\SI{0.2919}{}$ & $\SI{0.2795}{}$ & $\SI{0.2808}{}$\\
$c_p$ & $\SI{-2.995}{\femto\newton}$ & $\SI{-6.455}{\femto\newton}$ & $\SI{-2.268}{\femto\newton}$ & $\SI{-6.007}{\femto\newton}$\\
$a_v$ & $\SI{0.2506}{\femto\newton}$ & $\SI{0.2033}{\femto\newton}$ & $\SI{0.3568}{\femto\newton}$ & $\SI{0.2686}{\femto\newton}$\\
$b_v$ & $\SI{0.2447}{}$ & $\SI{0.2180}{}$ & $\SI{0.2517}{}$ & $\SI{0.2104}{}$ \\
$c_v$ & $\SI{3.073}{\femto\newton}$ & $\SI{6.943}{\femto\newton}$ & $\SI{2.359}{\femto\newton}$ & $\SI{6.529}{\femto\newton}$
\end{tabular} 
\end{ruledtabular}%
\caption{\label{tab:Fitparameter}Fit parameters of the function \eqref{eq:Fit} for the force components $F_{\mathrm{p},p}$ (index $p$) and $F_{\mathrm{p},v}$ (index $v$) for each considered particle shape. The second-last column corresponds to Fig.\ \ref{fig:Fit_Plot}.}
\end{table*}

\section{\label{results}Results and discussion}
\begin{figure}[htb]
\centering
\includegraphics[width=\linewidth]{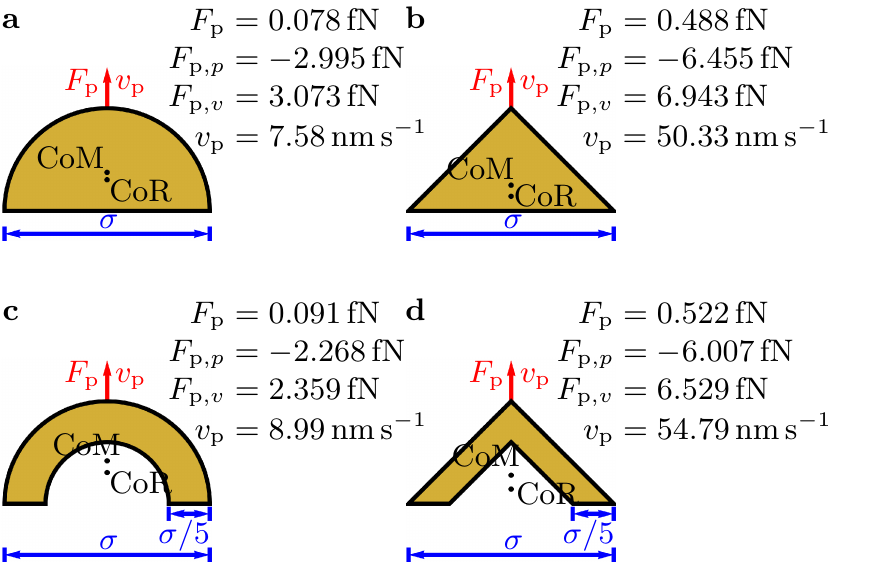}%
\caption{\label{fig:particle_shapes}The considered particle shapes: (a) half ball, (b) cone, (c) hollow half ball, and (d) hollow cone. All particles have diameter $\sigma$ and the hollow particles have wall width $\sigma/5$. The center of mass (CoM), the center of resistance (CoR), and the direction of the propulsion force $F_{\mathrm{p}}$ and corresponding propulsion speed $v_{\mathrm{p}}$, which are parallel to the symmetry axes of the particles and perpendicular to the main direction of sound propagation, are indicated. For each particle shape, the values of $F_{\mathrm{p}}=F_{\mathrm{p},p}+F_{\mathrm{p},v}$, the pressure component $F_{\mathrm{p},p}$, the viscous component $F_{\mathrm{p},v}$, and $v_{\mathrm{p}}$ are given.}
\end{figure}

Since we are interested in studying the effect of pointedness and hollowness on the particle propulsion, we consider four different particle shapes: a half ball, a cone, and hollowed-out versions of both shapes (see Fig.\ \ref{fig:particle_shapes}).  
All considered particles have diameter $\sigma$ and the hollow ones have wall width $\sigma/5$. 
Both the particles' center of mass (CoM) and center of resistance (CoR) are on the symmetry axes of the particles. 
These particle shapes are chosen, since they are relatively simple with an axis of rotational symmetry, have a head-tail asymmetry that is necessary for acoustic propulsion, and differ with respect to pointedness and hollowness. The half ball is neither pointed nor hollow, the cone is pointed but not hollow, the hollow half ball is not pointed but hollow, and the hollow cone is both pointed and hollow. Due to the huge computational expense of the simulations, we did not consider additional particle shapes. A further motivation for choosing the mentioned particle shapes is the fact that there exist experimental data from a previous study that considered cup-shaped particles that are similar to our hollow half ball \cite{SotoWGGGLKACW2016}. 

Figure \ref{fig:particle_shapes} shows also the results for the propulsion force $F_{\mathrm{p}}=F_{\mathrm{p},p}+F_{\mathrm{p},v}$, its pressure component $F_{\mathrm{p},p}$ and viscous component $F_{\mathrm{p},v}$, and the corresponding propulsion speed $v_{\mathrm{p}}$.
For all particle shapes, $F_{\mathrm{p}}$ and $v_{\mathrm{p}}$ are positive. The components $F_{\mathrm{p},p}$ and $F_{\mathrm{p},v}$ are always negative and positive, respectively, where the latter component is dominating. Remarkably, both cone-shaped particles are associated with propulsion speeds $v_{\mathrm{p}}\approx\SI{50}{\nano\meter\,\second^{-1}}$ that are one order of magnitude larger than those for the half-ball shapes. This suggests that pointed shapes allow much faster acoustic propulsion than rounded ones. Comparing the corresponding filled and hollowed-out particle shapes reveals that the hollow particles reach slightly (less than $20$ percent) larger propulsion speeds than their filled counterparts. This suggests that cavities in the particles have no significant effect on their propulsion speed. Among the considered particles, that with a hollow-cone shape reaches the largest propulsion force $F_{\mathrm{p}}=\SI{0.522}{\femto\newton}$ and the largest propulsion speed $v_{\mathrm{p}}=\SI{54.79}{\nano\meter\,\second^{-1}}$.

Our result for $v_{\mathrm{p}}$ for the hollow-half-ball particle can be compared to results from experiments described in Ref.\ \onlinecite{SotoWGGGLKACW2016}, where particles with a similar shape and size were found to propel with speed $v_{\mathrm{p}}=\SI{82.4\pm5.4}{\micro\metre\,\second^{-1}}$. However, the comparison is complicated by the fact that this reference mentions not the acoustic energy density the particles were exposed to, but instead, as it is usual in experimental studies on acoustically propelled particles, only the amplitude of the alternating voltage applied to the piezoelectric transducer. To estimate the energy density that is related to the known voltage amplitude, we use the typical energy-density values for some voltage ranges given in Ref.\ \onlinecite{Bruus2012}. According to this reference, a voltage amplitude lower than $\SI{10}{\volt}$, as is used in the experiments described in Ref.\ \onlinecite{SotoWGGGLKACW2016}, is typically associated with an acoustic energy density of $10$-$\SI{100}{\joule\,\meter^{-3}}$.  
In our simulations, the energy density was $\SI{22.7}{\milli\joule\,\meter^{-3}}$ and thus much smaller than in the experiments. Calculating the propulsion force $F_{\mathrm{p}}$ that corresponds to the propulsion speed $v_{\mathrm{p}}$ reported in the experiments and assuming that this force scales linearly with the acoustic energy density, we find a range of force values $[F^{\mathrm{exp}}_{\mathrm{min}},F^{\mathrm{exp}}_{\mathrm{max}}]$ with $F^{\mathrm{exp}}_{\mathrm{min}}=\SI{0.163}{\femto\newton}$ and $F^{\mathrm{exp}}_{\mathrm{max}}=\SI{1.63}{\femto\newton}$ that could have been observed in the experiments when using the same energy density as in our simulations. 
This is consistent with our finding for $F_{\mathrm{p}}$. To be precise, our value for $F_{\mathrm{p}}$ is slightly below $F^{\mathrm{exp}}_{\mathrm{min}}$, but given that $F^{\mathrm{exp}}_{\mathrm{min}}$ and $F^{\mathrm{exp}}_{\mathrm{max}}$ have been determined by a rough estimate, that we simulated traveling ultrasound waves whereas the experiments involved standing waves, and that the frequency of the ultrasound was different in the simulations and experiments, the agreement of the force interval estimated from the experimental data with our result is very good. 

\begin{figure*}[tb]
\centering
\includegraphics[width=\linewidth]{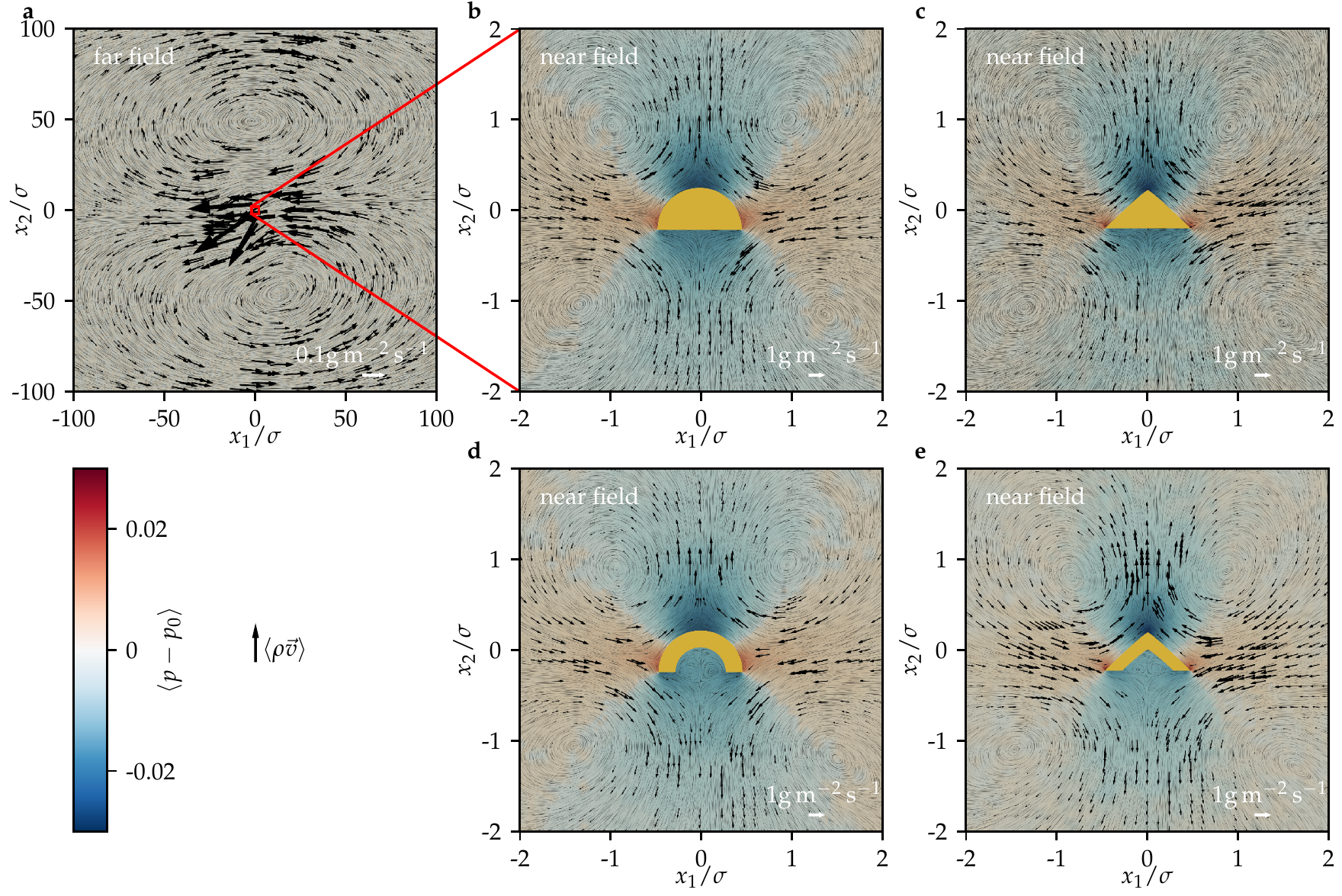}%
\caption{\label{fig:flowfield}The time-averaged mass-current density $\langle\rho\,\vec{v}\rangle$ and reduced pressure $\langle p-p_{0}\rangle$ for all considered particle shapes (see Fig.\ \ref{fig:particle_shapes}). (a) The far field is shown for a half-ball particle and looks similar for the other particle shapes; (b)-(e) The near field is shown for each particle shape.}
\end{figure*}

Next, we study the flow field around the particles. Figure \ref{fig:flowfield} shows the time-averaged mass-current density $\langle\rho\,\vec{v}\rangle$ and reduced pressure $\langle p-p_{0}\rangle$ for all considered particles. 
In each case, the far field of the flow shows two large counter-rotating vortices with diameters of about $100\sigma$, which are in front of and behind the particle, respectively.
Such a flow field is typical for acoustic streaming around a particle exposed to ultrasound even for simple shapes like a sphere \cite{XieEtAl2015}.
Also the near field of the flow is similar for all considered particle shapes. 
The structure of the near field, however, is qualitatively different from that of the far field. There are now four vortices to the left and right in front of and behind the particles. The diameter of these vortices is about $2\sigma$. 
There is an inflow towards the particles from their left and right and an outflow away from the particles at their front and back. 
For the reduced pressure, negative values are observed in front of and behind the particles, whereas positive values are observed to their left and right. 

Such flow fields are well-known from squirmers \cite{LaugaP2009,ShenWL2018}. Comparing the flow near fields observed in our simulations with those from different types of squirmers, our particles can be identified as \textit{pushers}. 
The identification of the acoustically-propelled particles as pushers is an important discovery, since it allows to translate the known properties of and modeling techniques for pushers to the so far much less investigated acoustically-propelled particles. 
Our results for the flow fields suggest that they are caused by local acoustic streaming close to the particles, as predicted in the theoretical work \onlinecite{NadalL2014}.

\section{\label{conclusions}Conclusions}
Based on direct numerical simulations, we have studied the acoustic propulsion of nonspherical nano- and microparticles by traveling ultrasound.
For particle shapes that were either rounded or pointed and either filled or hollow, we have calculated the propulsion force acting on the particle, the resulting propulsion speed, as well as the flow field around the particle. 
This allowed us to obtain important new insights into the propulsion of such particles. 
Our results, which are consistent with experiments from Soto \textit{et al.} \cite{SotoWGGGLKACW2016} for cup-shaped particles, confirmed that the particles' propulsion is very sensitive to the particle shape \cite{CollisCS2017}. The results revealed that a particle with a pointed shape can show a much more efficient propulsion than one with a rounded shape and that a cavity in the particle shape has no significant effect on the propulsion efficiency. 
Considering the small number of different particle shapes that have been addressed in previous studies, these findings suggest to use conical particles in future studies further investigating or applying acoustically-propelled colloidal particles to reach higher propulsion speeds than for cup-shaped particles \cite{SotoWGGGLKACW2016} and the commonly used bullet-shaped particles \cite{WangCHM2012,AhmedEtAl2013,GarciaGradillaEtAl2013,AhmedWBGHM2016}. Since we found only a negligible effect of a cavity in the particle shape on its propulsion efficiency, the conical particle can have or have not a cavity, depending on which particle shape is easier to synthesize. 
Knowing that the propulsion efficiency of the particles can significantly be enhanced by choosing a more suitable particle shape is an important insight with respect to potential applications of this type of active particles, e.g., in nanomedicine, where the particles need a large propulsion speed to withstand blood flow while the ultrasound intensity is limited to physiologically harmless values \cite{NitschkeW2020,CeylanYKHS2019,ErkocYCYAS2019}. When the particles shall be used for drug delivery in nanomedicine, a filled particle is advantageous, as its larger volume is associated with a larger capacity for transporting drugs.  

The obtained time-averaged flow fields support the understanding of the particles' propulsion mechanism as originating basically from local acoustic streaming, as predicted by Nadal and Lauga \cite{NadalL2014}. Remarkably, the flow's near field allowed us to identify the particles as pusher squirmers, which strongly extends our understanding of ultrasound-propelled nano- and microparticles.  
This finding allows to transfer the large existing knowledge about the motion of pushers and, e.g., their interactions with obstacles and other particles to ultrasound-propelled particles, for which most of these issues have not yet been addressed. 
We expect that this will strongly boost the future theoretical investigation of ultrasound-propelled colloidal particles. 
For example, one could now use modified squirmer models to describe the motion of the particles on much larger time scales, which are no longer set by the high frequency of the ultrasound but instead by the rather small time-averaged flow velocities near the particles. 
This would reduce the computational effort to simulate the motion of the particles by several orders of magnitude and enable studies that were well-nigh impossible up to now. In addition, such modeling would even allow for the application of new analytical approaches, such as field theories for describing the collective dynamics of many interacting particles on mesoscopic or macroscopic scales \cite{WittkowskiL2011,TiribocchiWMC2015,StenhammarWMC2016,WittkowskiSC2017,BickmannW2019}.  
Furthermore, the identification of the ultrasound-propelled particles as pushers has intriguing consequences for future materials science. Since it is known from suspensions of bacteria that pushers can strongly reduce the viscosity of a liquid, we can predict that it is possible to use suspensions of ultrasound-propelled particles to realize novel active materials with a viscosity that can be tuned via the ultrasound intensity from the normal positive viscosity of the suspension in the absence of ultrasound through to suprafluidity up to even negative viscosities \cite{Saintillan2018,LopezGDAC2015}.

Apart from that, the basic understanding of the details of the particles' propulsion mechanism should further be extended by additional computational fluid dynamics simulations. As a task for the future, e.g., the dependence of the propulsion efficiency on the aspect ratio of conical particles should be studied in detail. Examples for other parameters, whose influence on the particles' propulsion still needs to be studied, are the viscosity of the fluid and the frequency of the ultrasound.

\begin{acknowledgments}
We thank Patrick Kurzeja for helpful discussions. 
R.W.\ is funded by the Deutsche Forschungsgemeinschaft (DFG, German Research Foundation) -- WI 4170/3-1. 
The simulations for this work were performed on the computer cluster PALMA II of the University of M\"unster. 
\end{acknowledgments}

\nocite{apsrev41Control}
\bibliographystyle{apsrev4-1}
\bibliography{control,refs}
	
\end{document}